\documentclass[%
 aip,
 apl,
 amsmath,amssymb,
 reprint,%
]{revtex4-1}

\usepackage{graphicx, import}
\usepackage{dcolumn}
\usepackage{bm}
\usepackage{color}
\usepackage{soul}
\usepackage{float}

\usepackage[utf8]{inputenc}
\usepackage[T1]{fontenc}

\usepackage{tikz}
\usepackage{xcolor}

\definecolor{lightgrey}{rgb}{0.88,0.88,0.88}
\definecolor{lightred}{rgb}{1,0.85,0.88}

\newcommand{\pos}[1]{\mathbf{#1}}

\begin{document}


\title{Adversarial reverse mapping of condensed-phase molecular
structures: Chemical transferability}


\author{Marc Stieffenhofer}
\affiliation{%
	Max Planck Institute for Polymer Research, 55128 Mainz, Germany
}%


\author{Tristan Bereau}
\affiliation{
    Van 't Hoff Institute for Molecular Sciences and
    Informatics Institute, University of Amsterdam, Amsterdam 1098 XH, The
    Netherlands
}
\affiliation{%
  Max Planck Institute for Polymer Research, 55128 Mainz, Germany
}%

\author{Michael Wand}
\affiliation{
	Institute of Informatics, Johannes Gutenberg University, 55099 Mainz, Germany
}%


\date{\today}

\begin{abstract}
Switching between different levels of resolution is essential for
multiscale modeling, but restoring details at higher resolution
remains challenging. In our previous study we have introduced
deepBackmap: a deep neural-network-based approach to reverse-map
equilibrated molecular structures for condensed-phase systems. Our
method combines data-driven and physics-based aspects, leading to
high-quality reconstructed structures. In this work, we expand the
scope of our model and examine its chemical transferability. To this
end, we train deepBackmap solely on homogeneous molecular liquids of
small molecules, and apply it to a more challenging polymer melt. We
augment the generator's objective with different force-field-based
terms as prior to regularize the results. The best performing physical
prior depends on whether we train for a specific chemistry, or
transfer our model. Our local environment representation combined with
the sequential reconstruction of fine-grained structures help reach
transferability of the learned correlations.
\end{abstract}

\maketitle


\section{Introduction}
\label{sec:introduction}
The demand to further expand the accessible length and time scales in computer simulations for molecular systems remains consistently high. Breaking the limits of molecular dynamics (MD) simulations is therefore still an area of active research. State-of-the-art approaches include enhanced-sampling
techniques \cite{mitsutake2013enhanced}, dedicated hardware~\cite{shaw2009millisecond} and hierarchical multiscale modeling.\cite{kremer2002multiscale,
horstemeyer2009multiscale, peter2009multiscale}

Multiscale modeling aims at linking different levels of resolution.
The reduced resolution in a coarse-grained (CG) model smoothens the energy
landscape and thereby effectively accelerates the simulation. On the
other hand, atomistic details are sometimes necessary for a thorough
investigation of processes on smaller scales. The goal is
therefore to use a CG model with a reduced number of
degrees of freedom where it is possible and switch back to a higher
resolution where it is needed.\cite{praprotnik2008multiscale, zhang2019hierarchical}
However, the process of reintroducing lost degrees of freedom is challenging as it requires to reinsert details with the correct statistical weight: Given the CG configuration the generated atomistic structure should follow the Boltzmann distribution of atomistic microstates.

Existing backmapping schemes typically consist of the following steps: At
first, an initial atomistic structure is proposed for the given CG
configuration.\cite{tschop1998simulation2} A generic approach for this is to randomly place the
atoms close to their corresponding CG bead center.\cite{rzepiela2010reconstruction,wassenaar} Subsequent energy
minimization is needed to relax the structures and some (typically
position restrained) MD simulations have to be performed to obtain the
correct Boltzmann distribution. The computational effort for the
energy minimization and MD simulation schemes can become significant.
Further, poorly initialized structures can get trapped into local
minima with high energy barriers. Therefore human intervention is still
required for more complex molecular structures to obtain a reasonable
initial structure.
The computational cost of subsequent energy minimization and MD
simulation procedures can be reduced significantly when presampled
fragments of a correctly sampled all-atom structures are used to
generate the initial backmapped
configuration.\cite{hess2006long,peter2009multiscale,
zhang2019hierarchical, brasiello2012multiscale} 

While reverse-mapping of molecular structures is still tackled largely by classical methods like energy minimization and MD simulation, recent approaches leveraging machine learning (ML) are receiving growing attention. Wang and G\'omez-Bombarelli used a variational auto-encoder (VAE) to learn a mapping from an all-atom representation to coarse-grained variables, parametrizing the coarse-grained force field and decoding back to atomistic detail.\cite{wang2019coarse} 
Other approaches, including our previous study and the work by Li \emph{et al.}, used convolutional conditional generative adversarial networks (convolutional cGANs) to learn the correspondence between CG and fine-grained configurations.\cite{previous_work, li2020backmapping} cGANs originate from computer vision applications and have shown the ability to model highly complex and detailed probability distributions.\cite{pggan} Li \emph{et al.} used a convolutional cGAN for their study on backmapping cis-1,4 polyisoprene melts using an image representation by converting XYZ components of vectors into RGB values.\cite{li2020backmapping}
Other approaches for generating low-energy geometries for molecular compounds but not specifically designed for reverse-mapping include autoregressive models,\cite{Oord16,gebauer2018generating} invertible neural network,\cite{noe2019boltzmann} Euclidean distance matrices,\cite{hoffmann2019generating} and graph neural
networks.\cite{mansimov2019molecular}

In our previous work we have introduced deepBackmap (DBM):\cite{previous_work} An approach based on cGANs to directly predict equilibrated molecular structures for condensed-phase systems. In contrast to the work by Li \emph{et al.}, we aim to improve the quality of generated structures by incorporating prior knowledge into the input representation as well as the loss function of the generator. We use a voxel representation to encode spatial relationships and make use of different feature channels typical for convolutional neural networks to encode information of the molecular topology. The loss function of the generator is augmented with a term penalizing configurations with high potential energy. 

A regular discretization of 3D space prohibits scaling to larger spatial structures. Therefore, we use an autoregressive approach that reconstructs the fine-grained structure incrementally, atom by atom. In each step, we provide the convolutional generator only with local information, making the method scalable to larger system sizes and applicable to condensed phase systems.

In this work we explore the model's capability with respect to
chemical transferability: we probe model generalization beyond the
chemistry used for training. We recycle the learned local correlations
to make predictions for molecules absent from the training set. We
argue that our sequential approach combined with the local-environment
representation is well suited to achieve chemical transferability, as
long as the generation of one atom only relies on short-range
force-field related features. We hypothesize that these atomic
environments strongly overlap across chemistry, as suggested by the
successes of ML for various electronic
properties.\cite{vonLilienfeld2020} We train the model on molecular
liquids of small molecules: octane and cumene. After training, we
deploy the model on a more challenging polymeric melt: syndiotactic
polystyrene (sPS). sPS is well suited for our study as it is
sufficiently complex but still has some features in common with octane
and cumene and therefore allows for a better understanding of the
limits of generalization. The pertinent but imperfect match between
the small molecules and polymer make for a more stringent backmapping
exercise. Furthermore, we insert two different physical priors into
the generator's objective based on the molecular force field. We
compare their impact on the performance of the model, especially
regarding chemical transferability.

\begin{figure*}[ht]
	\begin{center}
	    \includegraphics[width=0.8\linewidth]{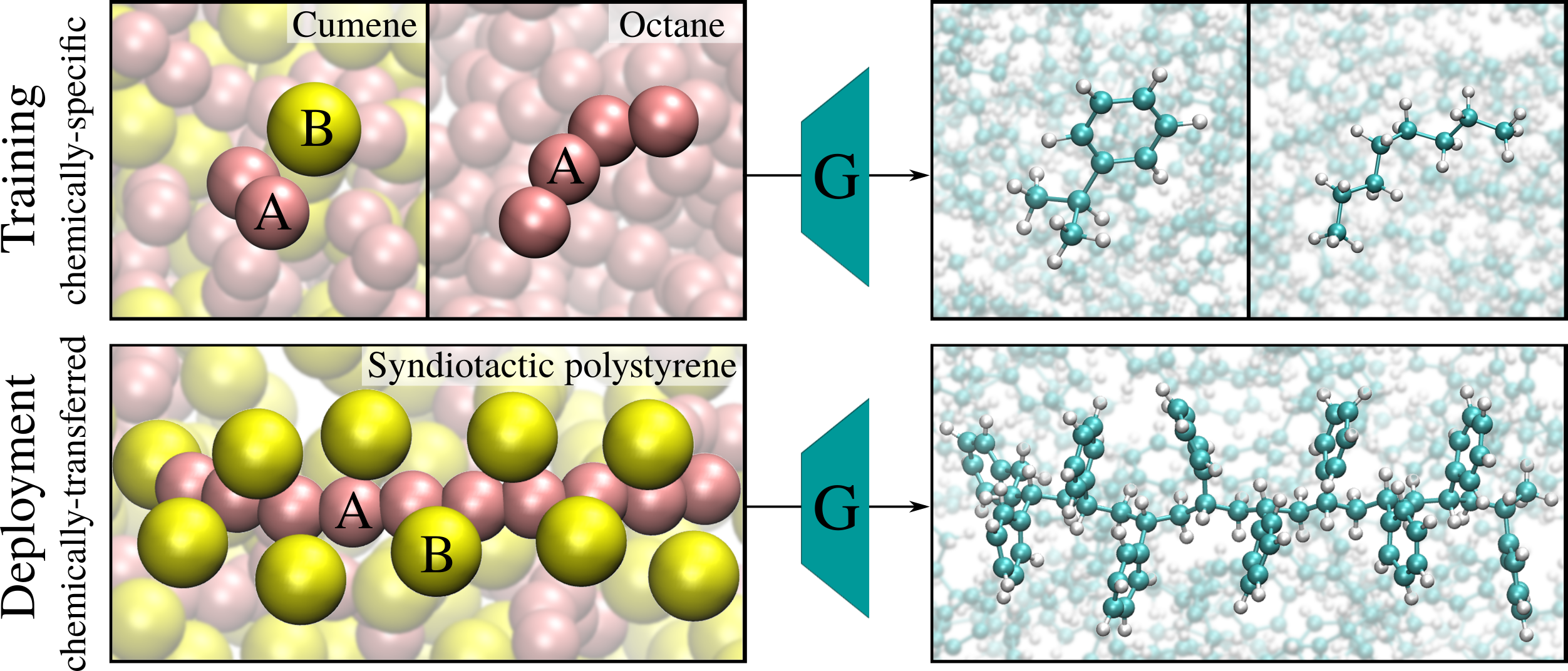}
	    \caption{All-atom and coarse-grained representations of
	    different molecules. A similar coarse-to-fine mapping is used
	    for all molecules (beads denoted A for the chain backbone and
	    B for the phenyl ring). We employ the chemical transferability
	    of our ML model DBM by training it solely on octane and cumene
	    structures and then apply it to the more challenging system of
	    syndiotactic polystyrene. }
	    \label{intro}
	\end{center}
\end{figure*}

\section{Machine learning model}
\label{sec:mlm}
In the following, we will briefly summarize the approach and then
focus on new extensions and applications of our model DBM. For a more
detailed description of the model the reader is referred to our recent
publication.\cite{previous_work}

\subsection{Setup}
We recall the notation for the coarse-grained and atomistic resolutions,
as well as the backmapping procedure: 
\begin{description}
	\item [Coarse-grained resolution] Let $\{{\bm A}_I = (\pos{R}_I, C_I) \vert
	I=1,\dots,N\}$ denote the set of $N$ coarse grained beads, where $I(i)$ is the index of the bead that contains atom with index $i$. Each bead
	has position $\pos{R}_I \in \mathbb{R}^3$ and bead type $C_I$.
	\item [Atomistic resolution] Let $\{{\bm a}_i = (\pos{r}_i, c_i) \vert
	i=1,\dots,n\}$ denote the set of $n$ atoms, with position $\pos{r}_i
	\in \mathbb{R}^3$ and atom type $c_i$. We denote $\varphi_I \subset
	\{{\bm a}_i | i=1,\dots,n\}$ as the set of atoms contained in the
	coarse-grained bead ${\bm A}_I$.
	\item [Backmapping] Backmapping requires us to generate a set of $n$
	atom positions $\pos{r}_1,\dots,\pos{r}_n$ conditional on the
	coarse-grained (CG) structure, given by the $N$ beads $A_1, \dots ,
	A_N$, as well as the atom types $c_1, \dots, c_n$. We express this
	problem as a conditional probability $p(\pos{r}_1,\dots,\pos{r}_n |
	c_1, \dots, c_n, {\bm A}_1, \dots,{\bm A}_N)$. 
\end{description}

Our ML technique takes examples
of corresponding coarse- and fine-grained configurations as input and from
this training data learns to generate further samples from the conditional distribution $p$.

Learning to sample from $p(\pos{r}_1,\dots,\pos{r}_n \vert c_1, \dots
, c_n, {\bm A}_1, \dots, {\bm A}_N)$ directly causes several problems:
(1) The trained model is fixed on the system size and atom ordering
used during training. (2) The model becomes specific for the given
molecules and thus chemical transferability can not be achieved. (3) A
dense system has an overwhelming number of degrees of freedom. A
one-shot method, which generates all coordinates of the system at
once, would have to solve an unreasonably high dimensional problem.
Applications to the condensed phase are therefore limited.

To avoid these problems, we factorize $p$ in terms of atomic contributions, where the
generation of one specific atom becomes conditional on CG beads
as well as all the atoms previously
reconstructed.\cite{gebauer2018generating} We therefore train a generative network, $G$, to generate and refine the
atom positions sequentially.

The backmapping scheme hereby consists of two steps: ($i$) An initial
structure is generated using the factorization
\begin{widetext}
	\begin{equation}
		\label{factorization}
		p(\pos{r}_1,\dots,\pos{r}_n \vert c_1, \dots , c_n, 
		{\bm A}_1, \dots, {\bm A}_N) 
		= \prod_{i = 1}^n p\left({\pos{r}}_{S(i)} \vert
		\pos{r}_{S(1)},\dots,\pos{r}_{S(i-1)},
		c_{S(1)},\dots,c_{S(i)}, 
		{\bm A}_1,\dots, {\bm A}_N\right), 
	\end{equation}
\end{widetext}
where $S$ sorts the atoms in the order of reconstruction and
$\{\pos{r}_{S(1)},\dots,\pos{r}_{S(i-1)}\}$ correspond to atoms that
have been already reconstructed. The dependence on earlier predictions
of $G$ makes our approach \emph{autoregressive}. In a second step, we want to refine the generated structures, since this approach can still lead to misplaced atoms that blow up the potential energy of the system.
To this end, we perform a sampling scheme inspired by Gibbs sampling, which iteratively resamples along the sequence $S$ several
times.\cite{GemanGeman84} Each further iteration still updates one
atom at a time, but uses the knowledge of \emph{all} other atoms.
Our experiments confirmed that such sampling leads to a good
approximation of $p$, even with a small number of iterations and
fixing the atom ordering.

\subsection{Local representation and feature embedding}

We use deep convolutional neural networks (deep CNNs), motivated by
the impressive developments for generative tasks in computer vision.\cite{fukushima1980neocognitron, pggan} In order to leverage CNNs for our task, an explicit spatial
discretization of ambient space is required. To this end, we use a
voxel-based representation. One-hot encoding of the atom positions, where each atom is assigned to its nearest voxel, leads to severe sparsity hindering learning. Therefore, we encode atoms and CG beads with a smooth
density, $\gamma({\bm x})$ and $\Gamma({\bm x})$, respectively,
modeled using Gaussian distributions

\begin{equation}
    \gamma_i ({\bm x}) = \exp \left( - \frac{({\bm x} - {\bm r}_i)^2}{2\sigma ^2} \right),
\end{equation}
where ${\bm x}$ is the spatial location in Cartesian coordinates expressed on a discretized grid. The density is centered around particle position ${\bm r}_i$ with Gaussian width $\sigma$, treated as a hyper parameter.

For each atom to be placed a unique representation is generated by
means of the density of particles placed around it. We assume locality
by limiting the amount of information about the environment to a
cutoff $r_\textup{cut}$ and sum over all atoms or beads within a cubic
environment of size $2 r_{\mathrm{cut}}$ centered on the current atom
of interest. In our previous work, we motivated the local environment
representation with the high computational costs of regular 3D grids (thus representing the whole structure at once gets infeasible) and the scalability to larger system sizes. In this study, we also
emphasize the gains of a local environment description for achieving
chemical transferability. It only encodes small-scale features that
are not necessarily unique for a given molecule and thus the learned
local correlations are more likely to generalize. 

Another key for the chemical transferability of DBM is the feature embedding. Similar to the three feature channels found for RGB images, we store a number of feature channels in each voxel that represent the presence of other atoms or beads of a certain kind. In our current implementation we made the feature mapping rather flexible such that it can be defined individually by the user. Atom types can be distinguished not only by element but additionally by chemical similarity, i.e., atoms of a given type can be treated as identical in the MD simulation. Furthermore, the user can add channels to distinguish the functional form of interaction to the current atom of interest. Interaction types can include bond, bending angle, torsion, and Lennard-Jones. Similarly, separate channels can be used to encode the different coarse-grained bead types. This feature representation is permutationally invariant with respect to the ordering of the atoms in the local environment. It is well suited for achieving chemical transferability as it focuses on local geometries determined by the underlying force field, rather than features specific to the molecule.

\subsection{Generative model}
We train our model using the generative adversarial approach. GANs yield strong results when aiming at high-quality generative tasks, in particular on high-dimensional and hard to model image spaces.\cite{pggan} The
generator, $G$, is trained against a second network, the critic, $C$.
While the critic, $C$, is trained to learn a distance metric between
generated and reference data, the generator, $G$, is trained to
minimize the distance. 

We use a cGAN to generate new atom positions. For atom $i$ contained in bead $I$ the input for $G$ is made up from a random noise vector $z \sim \mathcal{N}(0,1)$ and the conditional input $u_i := \{\xi_{i}, \Xi_{I(i)}, c_i \}$ consisting of the local environment representation for atoms $\xi_{i}$ and for beads $\Xi_{I(i)}$, as well as the current atom type $c_i$. The output of the generator $G$ is a smooth-density representation $\hat\gamma_i := G(z,u_i)$.

The critic network $C$ is trained to distinguish between reference densities $\gamma_i$ related to the conditional input $u_i$ and generated densities $\hat\gamma_i = G(u_i, z)$. We can write the basic loss function for the critic as
\begin{align}
    \mathcal{L}_C =\underset{i}{\mathbb{E}}
    \big[  &  C(u_i, \gamma_i) - C\left(u_i, G(u_i, z)\right)
    \big],
\end{align}
and for the generator we obtain a loss function purely affected by the generated data

\begin{align}
    \mathcal{L}_G =\underset{i}{\mathbb{E}}
    \big[  C\left(u_i, G(u_i, z)\right)
    \big].
\end{align}

\subsection{Extensions}
We reimplemented the model using the python package \textit{PyTorch}.\cite{paszke2019pytorch}
The code can be found at \url{https://github.com/mstieffe/deepBM}. In the
following we want to focus on the differences of the model compared to
our previous study.

\subsubsection{Regularization and normalization}

We use a variant of adversarial models where the Wasserstein distance, which arises from the idea of optimal transport,
serves as a metric to measure the similarity between the target and
the generated distributions.\cite{kolouri2017optimal} Computing the Wasserstein distance
directly is intractable, as it involves computing the infimum over the set of all possible joint probabilities of the target and generated distribution. Instead, we are able to use a dual representation of the Wasserstein distance, which is based on the Kantorovich-Rubinstein
duality, and use the critic $C$ to approximate it.\cite{wasserstein}
For this purpose, the critic $C$ has to be constrained to the set of 1-Lipschitz
functions. Two major approaches for achieving this are regularization
and normalization.

\textbf{Gradient Penalty}
A differentiable function is 1-Lipschitz if and only if it has gradients everywhere with norm at most one. A soft version of this constraint is enforced with a penalty on the gradient norm\cite{NIPS2017_7159}
\begin{align}
    \mathcal{L}_C =\underset{i}{\mathbb{E}}
    \big[  &  C(u_i, \gamma_i) - C\left(u_i, G(u_i, z)\right) \nonumber \\
    &+ \lambda_\textup{gp} \left(\lVert\nabla _{\tilde{u}_i,\tilde\gamma_i}
    C(\tilde{u}_i,\tilde\gamma_i)\rVert_2-1\right)^2
    \big],
\end{align}
where $(\tilde u_i, \tilde{\gamma}_i)$ is interpolated linearly between pairs of points $(u_i, \gamma_i)$ and $(u_i, G(u_i, z))$. The prefactor $\lambda_\textup{gp}$ scales the weight of the gradient penalty. The additional term in the loss function may be considered as a regularizer for the complexity of the critic $C$.\cite{}

\textbf{Spectral Normalization}
The Lipschitz constant of a linear function is its largest singular value (spectral norm). The 1-Lipschitz constrain can therefore be achieved by applying Spectral Normalization to all the weights in the network
\begin{equation}
	W \rightarrow \frac{W}{\sigma{(W)}},
\end{equation}
where $\sigma{(W)}$ is the largest singular value of $W$.\cite{miyato2018spectral}

In our previous study we used only regularization but in this study we found that combining regularization and normalization lead to the best results.

\subsubsection{Physical prior}

We collapse the generated smooth-density representation $\hat\gamma_i$ back to point coordinates by computing a weighted average, discretized over the voxel grid
\begin{equation}
 \hat{\bm r}_i = \int {\rm d}{\bm x}\, \hat\gamma_i({\bm x}) 
 \approx \sum_m \sum_k \sum_l x_{mkl}\hat\gamma_i(x_{mkl}).
\end{equation}
This density-collapse step is differentiable and thus the point
coordinates can be used to incorporate a physical prior, $p$, in the
loss function for the generator. $p$ is built on force-field-based
energy contributions and penalizes high-energy structures. It thereby effectively narrows down the functional space of the generator. Adding $p$ with appropriately low weight to the loss function helps steering the optimization and regularizes the generator. It aims at improving generalization and accelerating convergence. $p$ depends on the set of atoms
corresponding to a coarse-grained bead, $\varphi_I$ for reference
atoms and $\hat{\varphi}_I$ for generated atoms, as well as reference
atoms $N_I$ in the local neighborhood of different beads. In the
following, $\varepsilon_{t}$ refers to the potential energy of
specific intra- and intermolecular interactions, where $t$ runs over
the interaction types: intramolecular bond, angle, and dihedral, and
non-bonded Lennard-Jones. While bonded interactions are expressed via harmonic (bond, angle, improper dihedral) or periodic (proper dihedral) potentials, non-bonded interactions follow the Lennard-Jones potential 
\begin{equation}
 V_\textup{LJ}(r) = 4 \epsilon \left[ \left( \frac{\sigma}{r} \right)^{12} - \left( \frac{\sigma}{r} \right)^6 \right],
\end{equation}
where $\epsilon$ is the depth of the potential well and $\sigma$ its
characteristic distance.
In this study we compare two different prior types:

\textbf{Energy minimizing} The first prior $p_1$ aims at minimizing the potential energy of generated structures
\begin{equation}
    p_1(\varphi_I, \hat{\varphi}_I, N_I)
    = \sum_t \lambda_t \varepsilon_{t}(\hat{\varphi}_I, N_I).
\end{equation}

\textbf{Energy matching} The second prior $p_2$ penalizes \emph{discrepancies} between the potential energies of generated and reference structures
\begin{equation}
    p_2(\varphi_I, \hat{\varphi}_I, N_I)
    = \sum_t \lambda_t \left|\varepsilon_{t}(\varphi_I, N_I) - 
    \varepsilon_{t}(\hat{\varphi}_I, N_I)\right|.
\end{equation}
The prefactor $\lambda_t$ scales the weight of a given interaction term. 
Overall we use the following loss function for the generator
\begin{equation}
    \mathcal{L}_G = \underset{I}{\mathbb{E}}
    \Big[  
        \underset{i \in \theta_I}{\mathbb{E}} 
	    \big[
	      C\left(u_i, G(u_i, z)\right) 
	    \big]
        + p(\varphi_I, \hat{\varphi}_I, N_I)
    \Big],
\end{equation}
where $\theta_I = \{i | a_i \in \varphi_I\}$ is the set of atom indices for atoms contained in $\varphi_I$ and $p$ is one of the prior terms defined above.

\subsection{Implementation details}

We use 3D convolutional neural networks (CNNs) with
residual connections similar to our previous work.\cite{he2016deep} 

The model is trained for  60 epochs in total using a
batch size of 64. We start training with $\lambda_t =
0$ and increase it in small increments to $\lambda_t = 4 \cdot 10^{-2}$ for non-bonded Lennard Jones and $\lambda_t = 4 \cdot 10^{-3}$ for bonded interactions during training. The final values for $\lambda_t$ are obtained in a hyper parameter search optimizing the overall performance of the model. Treating all interaction terms equally (e.g. setting $\lambda_t = 4 \cdot 10^{-2}$ for all terms) lead to only marginal improvement regarding covalent interactions but significantly higher Lennard-Jones energies. On the other hand, setting $\lambda_t = 0$ for all covalent terms but keeping $\lambda_t = 4 \cdot 10^{-2}$ for Lennard-Jones makes the training unstable.
We use the Adam optimizer with learning rates $5 \cdot
10^{-5}$ for the generator and $10^{-4}$ for the critic. The prefactor scaling
the weight of the gradient penalty term is set to $\lambda_\textup{gp} = 0.1$. 
The critic $C$ is trained five times in each
iteration while the generator $G$ is trained just once.

As in our previous work, we train the model recurrently on atom
sequences containing either all heavy (carbon) or light (hydrogen)
atoms corresponding to a single coarse-grained bead. While CG force
fields might lead to the sharing of an atom between two neighboring
beads (see Sec.~\ref{sec:exp}), the reconstruction of the atom is
assigned to only one of the two beads. This assignment has no impact
on the local environment representation of the atoms (except a shift
of the center), but might affect the order of reconstruction. For
heavy atoms we remove intramolecular hydrogens from the local
environment representation. In training mode, the initial local
neighborhood for a sequence is generated from training data. After
each step, the generated atom density is added to the local
environment representation for the next atom in the sequence until all
atoms of the sequence are generated. In evaluation mode, no training
data is used and all environment atoms are generated autoregressively.

We reduce the rotational degrees of freedom by aligning the local environment according
to the position of the central bead and the difference vector to a bonded bead. This leaves one rotational degree of freedom around the director axis, for which we augment the training set by means of rotations. To further improve the quality of reconstructed structures, we
feed different orientations about said axis during prediction and choose the structure with the
lowest energy from the generated ensemble. We use four iterations to refine the structures.

\section{Computational methods}
\label{sec:exp}
This study is based on two molecular liquids: octane and cumene, as
well as a syndiotactic polystyrene (sPS) melt. All data were generated
using the molecular dynamics package {\sc GROMACS} (version 4.6 for
sPS, 5.0 for octane and cumene, but the version does not effect the outcome of the simulations).\cite{hess2008gromacs} Molecular
dynamics simulations were performed in the \emph{NPT} ensemble using
the velocity rescaling thermostat and the Parrinello-Rahman barostat.
An integration timestep of 1~fs was used.

The atomistic data for sPS was reported in Liu \emph{et
al.};\cite{liu2018polymorphism} the underlying force field is based on
the work of Mueller-Plathe.\cite{muller1996local} Replica Exchange
MD simulation, a temperature-based enhanced-sampling technique, was
used to sample the system. Each snapshot contains 36 chains consisting
of 10 monomers. For additional details regarding the simulations the
reader is referred to the work of Liu \emph{et
al.}\cite{liu2018polymorphism} The coarse-grained model of sPS was
developed by Fritz \emph{et al.}\cite{fritz2009coarse} It represents a
polymer as a linear chain, where each monomer is mapped onto two CG
beads of different types, denoted A for the chain backbone and B for
the phenyl ring (see Fig.~\ref{intro}). The center of bead A is the center of mass of the $\text{CH}_2$ group and the two neighboring CH groups, which are weighted with half of their masses. Bead B is centered at the center of mass of the phenyl group. Bonds are created between the backbone beads A--A and
between backbone and phenyl-ring beads A--B.

The atomistic data for the liquids of octane and cumene were generated
using the Gromos force field and the topologies were generated by the
{\sc Automated Topology builder}.\cite{malde2011automated} Notably,
the Gromos and sPS force fields differ in parametrization strategies,
leading to evident inconsistencies for both intra and intermolecular
interactions. Octane and cumene simulation boxes contained 215 and 265
molecules, respectively. Both systems were sampled at $350$\,K.
Similar to the sPS mapping, cumene was mapped onto three CG beads: Two
beads of type A for the backbone, each containing a methyl group and
sharing the CH group connected to the phenyl ring, and one bead of
type B for the phenyl ring. Octane was mapped onto four beads of type
A (Fig.~\ref{intro}), where neighboring A beads share a $\text{CH}_2$
group.

\section{Results}
\label{sec:results}
In the present work we want to probe the chemical transferability of
DBM. To this end, we train the model solely on a dataset consisting of
small molecules, but validate it on a challenging polymer melt. The
training set contains 
3,225 and 2,120 molecules
of octane and
cumene in the liquid state, respectively, both simulated at $T=350$\,K. After
training was completed, we applied DBM to 720 chains (7200 monomers) of sPS melt at $T=568$\,K. The
temperature discrepancy for the test and training set arises from the
different boiling and melting points of the molecules, as we wish to
probe the model's chemical transferability in their liquid state.
However, as we have shown in our previous work, DBM is robust against
temperature changes: the learned local correlations are weakly
sensitive to temperature.\cite{previous_work} Furthermore, we want to
investigate the impact of different force field-based priors used
during training of the model. We evaluate and compare the performance
of different models regarding their ability to reproduce structural
and energetic features of the sPS reference atomistic configurations.

\subsection{Local structural and energetic features}

\begin{figure}[h]
	\begin{center}
		\includegraphics[width=1.0\linewidth]{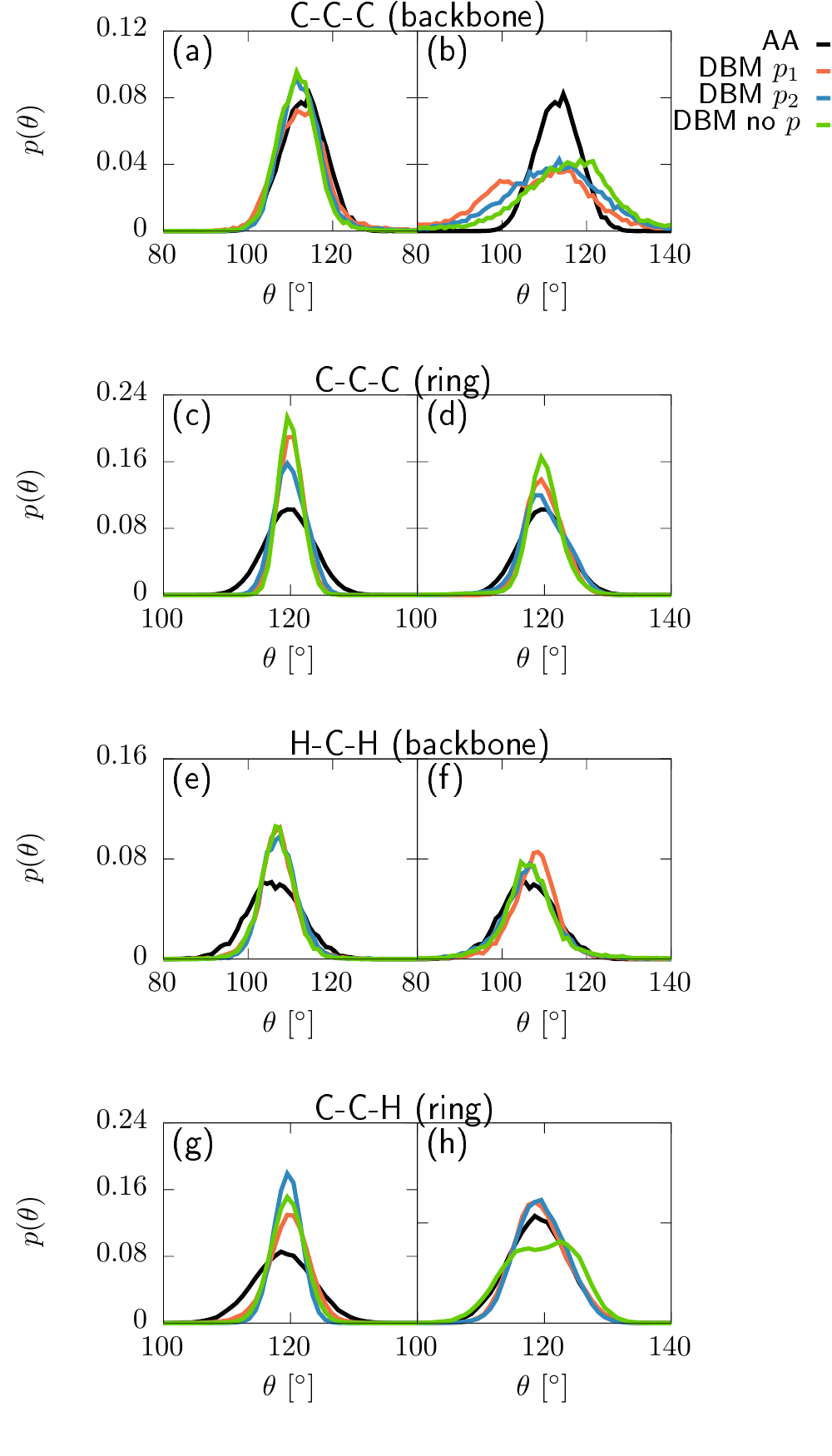}
		\caption{Canonical distributions for sPS at $T=568$\,K.
		Various angle terms for reference structures (``AA'') and
		structures generated with our method (``DBM'') using different
		priors during training are shown. Left: Chemically-specific model trained
		on sPS liquid at $T=568$\,K. Right: Chemically-transferred model trained on
		octane and cumene liquids at $T=350$\,K. }
		\label{results_angles} 
	\end{center}
\end{figure}

\begin{figure}[h]
	\begin{center}
		\includegraphics[width=1.0\linewidth]{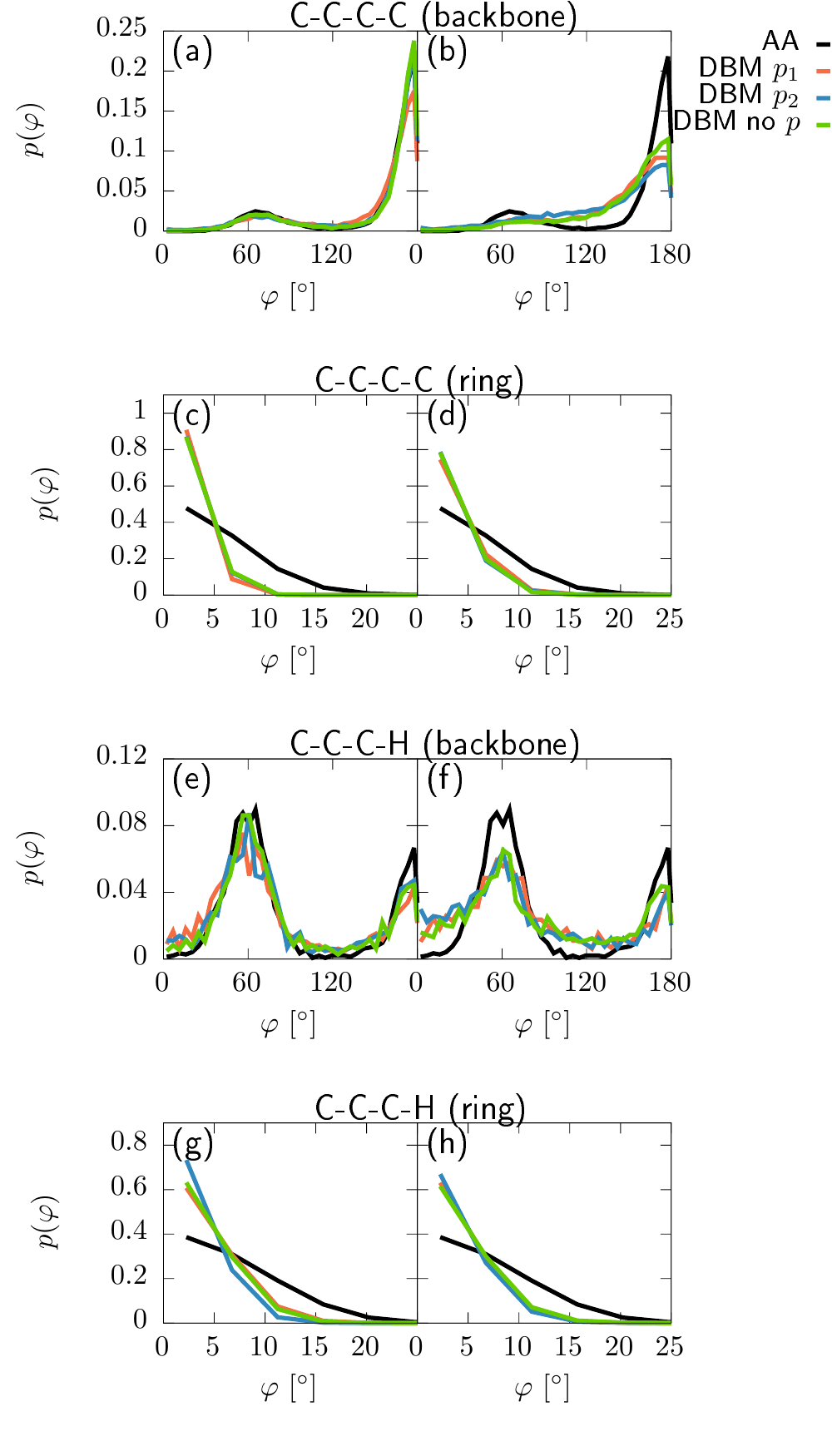}
		\caption{Canonical distributions for sPS at $T=568$\,K.
		Various dihedral terms for reference structures (``AA'') and
		structures generated with our method (``DBM'') using different
		priors during training are shown. (a,b,e,f) proper dihedral,
		(c,d,g,h) improper dihedral. Left: Chemically-specific model trained
		on sPS liquid at $T=568$\,K. Right: Chemically-transferred model trained on
		octane and cumene liquids at $T=350$\,K. }
		\label{results_dihs} 
	\end{center}
\end{figure}

\begin{figure}[htbp]
	\begin{center}
		\includegraphics[width=1.0\linewidth]{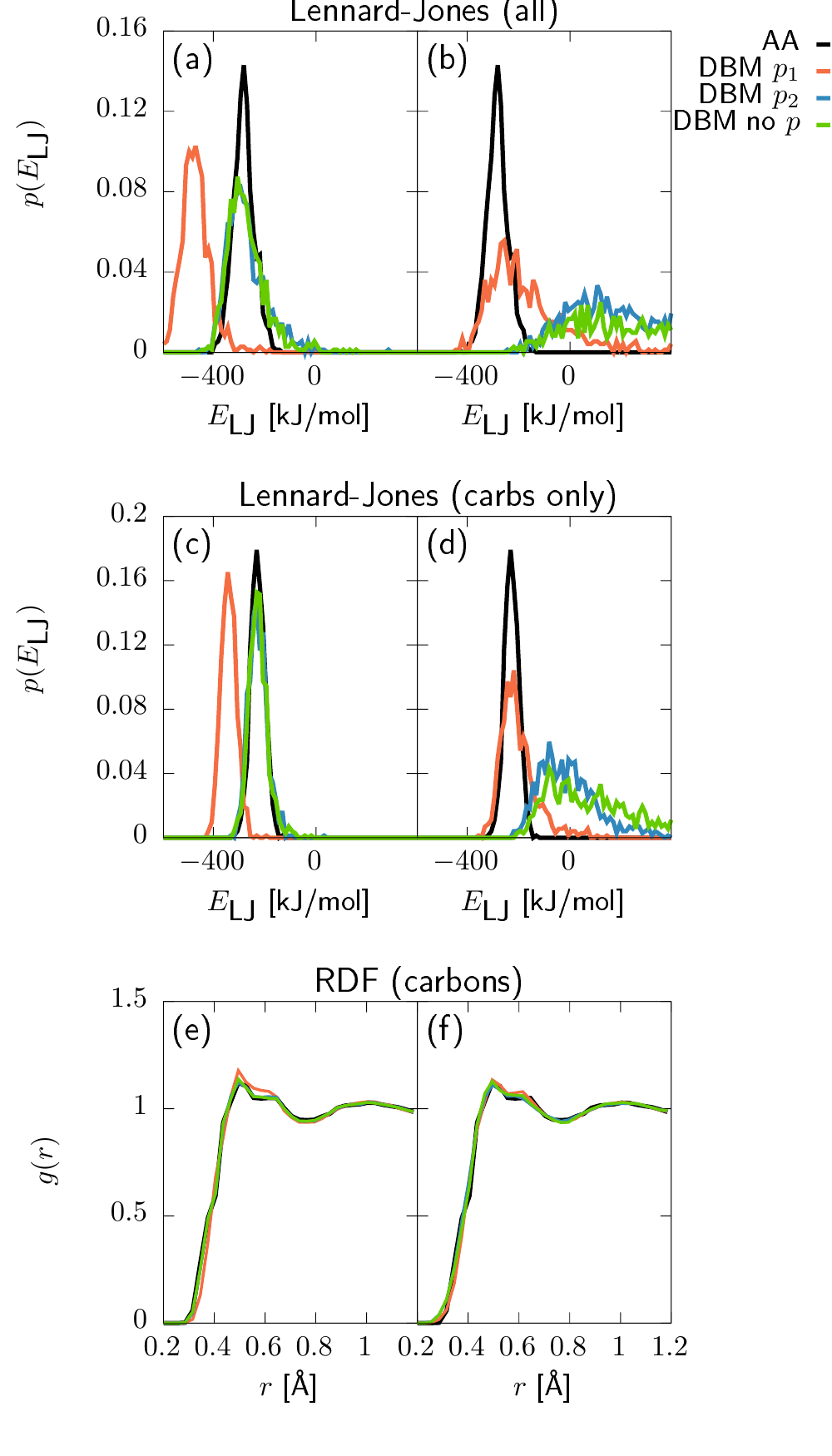}
		\caption{Canonical distributions for sPS at $T=568$\,K.
			Lennard-Jones energies for all atoms (a,b), Lennard-Jones
			energies for carbon atoms (c,d) and radial distribution
			functions, $g(r)$, of the non-bonded carbon atoms (e,f)
			for reference structures (``AA'') and structures generated
			with our method (``DBM'') using different priors during
			training are shown. Left: Chemically-specific model trained
		on sPS liquid at $T=568$\,K. Right: Chemically-transferred model trained on
		octane and cumene liquids at $T=350$\,K. }
		\label{results_non_bonded} 
	\end{center}
\end{figure}

Fig.~\ref{results_angles}--\ref{results_non_bonded} show distribution
functions for structural and energetic properties of sPS reference
structures (``AA'') and structures generated with DBM. The model was trained using three different prior
configurations: Prior $p_1$ (``energy minimizing''), prior $p_2$
(``energy matching'') and no prior. For a thorough comparison, we show
results for the chemically-specific models, i.e., trained directly on
sPS (left) and chemically-transferred models,
i.e., trained solely on octane and cumene configurations (right).

We first analyze the angle distributions shown in
Fig.~\ref{results_angles}. The largest discrepancy between
chemically-specific and chemically-transferred models can be found in
the C--C--C backbone-angle distributions (panels a and b). While
models trained directly on sPS are able to reproduce the distribution
with high accuracy, models trained solely on octane and cumene lead to
overly broad distributions. Surprisingly, for the other angles shown
(panels c--h) the opposite holds: Models trained on sPS generate
slightly too narrow distributions compared to the reference
configurations, but distributions generated with models trained on
octane and cumene are remarkably close to the reference system.
Apparently, the prior type used during training seems to only have a
marginal effect on the angle distributions. 

Next, we focus on the dihedral distributions displayed in
Fig.~\ref{results_dihs}. Again, the largest discrepancy between models
trained on the different training sets can be found for the
distributions of the C--C--C--C backbone dihedral (panels a and b),
which is well reproduced by models trained on sPS directly but models
trained on octane and cumene fail to reproduce the height of the main
peak and are not able to reproduce the side peak. Similarly, the
performance for the C-C-C-H backbone dihedral (panels e and f) is also
slightly worse for models trained only on the octane and cumene
dataset. On the other hand, improper dihedrals of the phenyl rings
(panels c, d, g, and h) are reproduced virtually equally well for both
training sets. While the generated improper dihedrals are slightly too
narrow compared to the reference distributions, we emphasize the small
range of angles due to the imposed planarity of the ring. The prior
used does not influence the generated dihedral distributions.
Performance vary between scenarios and interactions, as reported by
the Jensen-Shannon divergence between reference and backmapped
distributions (see Table S1 in the SI). We did not observe clear
trends between the quality of reconstruction between bending angle and
dihedral.

The impact of the prior becomes most significant for the Lennard-Jones
energies, shown in Fig.~\ref{results_non_bonded} (a--d) obtained for
each sPS chain separately. Regarding models trained on sPS, both prior
$p_2$ and no prior lead to a good match with the reference
distribution. While the carbon-only Lennard-Jones energies match well,
the generated distributions show slightly large high-energy tails. On
the other hand, prior $p_1$ over-stabilizes the system leading to a
significant shift of the distributions toward lower energies. This is
reasonable, since $p_1$ aims at minimizing the energy of generated
structures during training, and therefore might not account for the
diversity of generated microstates. For models trained on octane and
cumene these findings turn around: While using either prior $p_2$ or
no prior lead to significantly high Lennard-Jones energies, prior
$p_1$ dramatically improves the performance of the model. Training the
model to minimize the energy seems to help learning more general
features that are better transferable across chemistry. On the other
hand, the energy-matching prior $p_2$ and the absence of prior (only
data-driven) encourage the model to reproduce the features found in
the training set, making it less transferable. This is especially
relevant in the context of possible force-field inconsistencies.

\subsection{Large-scale structural features}

Next, we evaluate large-scale structural features. A first impression
can be gained looking at the pair correlation function, $g(r)$, shown
in Fig.~\ref{results_non_bonded} (e--f), obtained for pairs of
non-bonded carbon atoms. All models reproduce the pair correlation
function remarkably well, indicating that the pair statistics of the
sPS chains is reproduced with great accuracy. It more broadly suggests
good agreement regarding local packing. 

Beyond pair statistics, we seek to probe and compare the accuracy of
the generated configurations of the different models at higher order.
The higher dimensionality prevents us from directly visualizing the
space, and we turn to dimensionality reduction instead. We build a
two-dimensional map representing proximity relationships between sPS
monomers and their environments. The pairwise distance between two
such environments is encoded using a similarity kernel based on the
many-body smooth overlap of atomic position (SOAP) representation.\cite{bartok2013representing}
We neglect hydrogens in the representation. Taking the
pairwise similarity for $N$ monomers into account, we derive an $N
\times N$ similarity matrix. We further apply Sketchmap to project
this high dimensional representation of conformational space onto a
two dimensional embedding.\cite{tribello2012using,
ceriotti2011simplifying}

\begin{figure}[htbp]
	\begin{center}
		\includegraphics[width=0.7\linewidth]{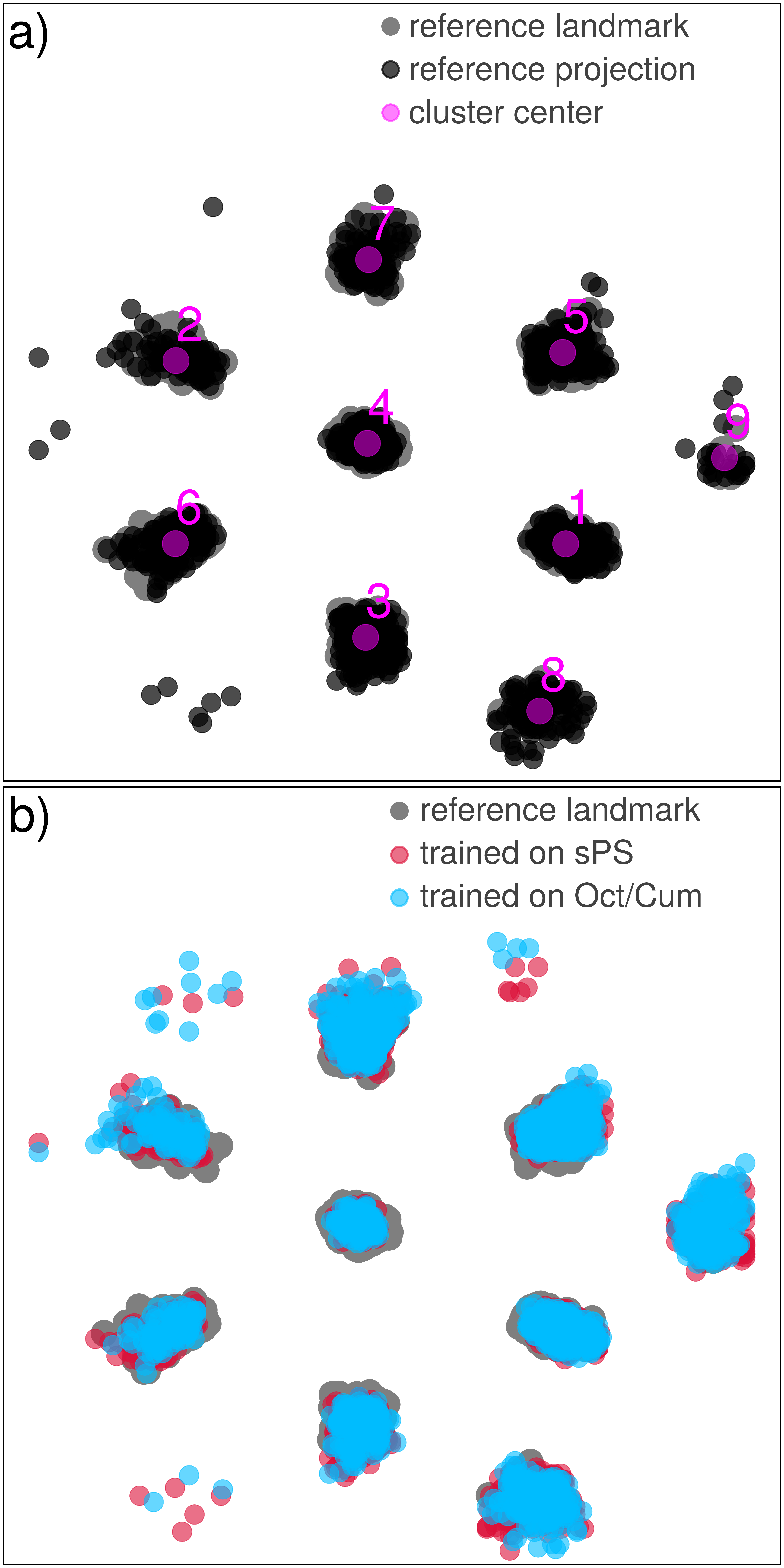}
		\caption{Low-dimensional structural space of condensed-phase
			configurations at $T=568$~K. For each panel, snapshots are
			backmapped from identical coarse-grained configurations,
			highlighting the overlap between reference and DBM structures.
			(a) landmarks and projections of reference structures and the
			obtained cluster centers. (b) landmarks of reference structures
			and projections of structures generated with DBM trained either
			on sPS or octane and cumene liquids trained with prior $p_1$.}
		\label{sm}
	\end{center}
\end{figure}

Fig.~\ref{sm} (a) displays a number of clusters obtained with the
Sketchmap algorithm using reference configurations. We analyze
the local environment of 720 sPS monomers to infer landmarks (gray) for
the two dimensional map. We used these landmarks to further project
1440 additional AA monomer representations (black). In Fig.~\ref{sm}
(b) we project the backmapped structures from corresponding
coarse-grained configurations applying a model trained on sPS (red) or
on octane and cumene (blue). Both models were trained using prior
$p_1$, the projections for models trained with different priors can be
found in Fig.~SI.7. The projections of structures
generated with DBM overlap significantly with the reference points,
indicating high structural fidelity beyond the pairwise level. 

\begin{figure*}[ht]
	\begin{center}
	    \includegraphics[width=0.95\linewidth]{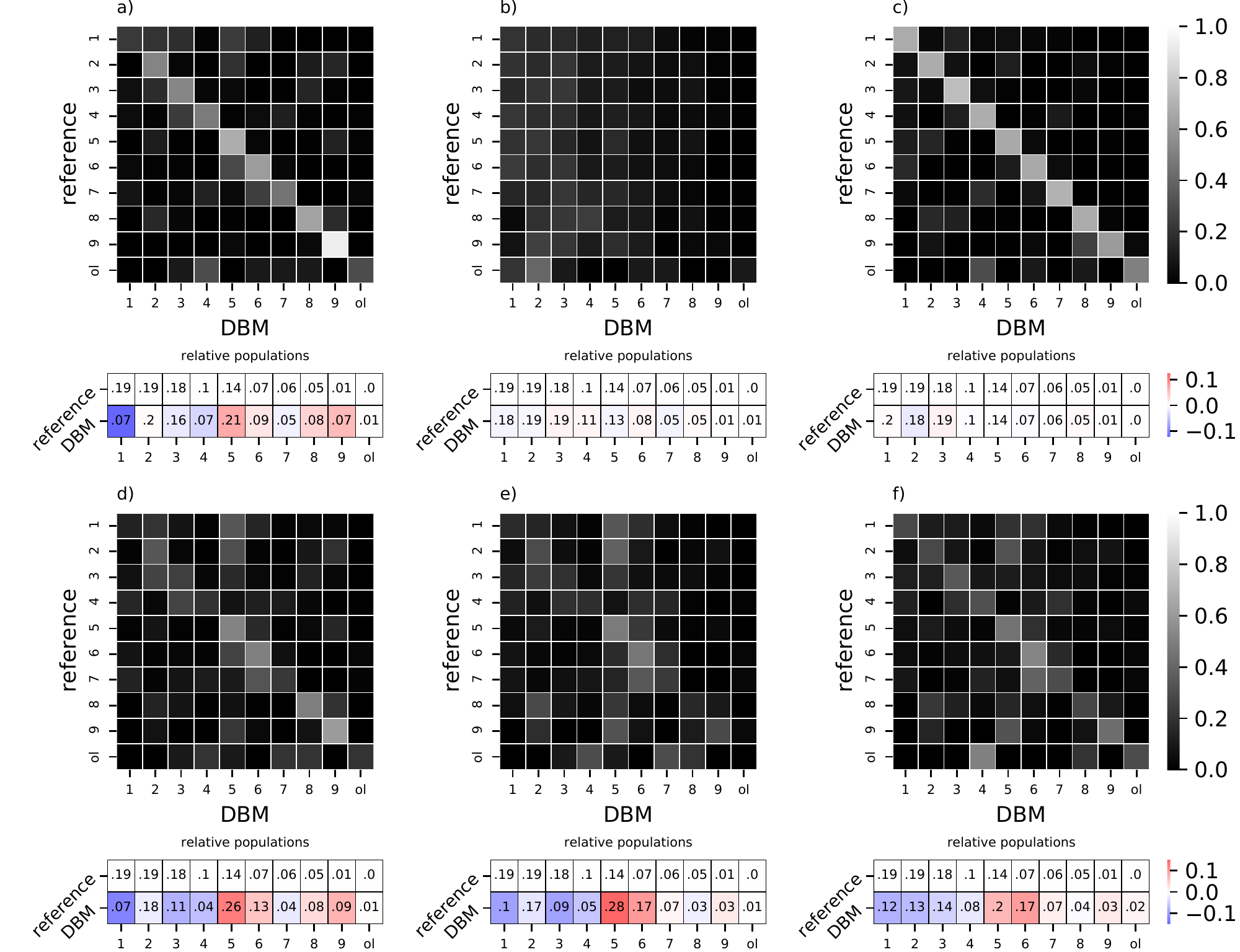}
	    \caption{(Top) Confusion matrix for the different clusters
	    obtained in the two-dimensional Sketchmap. (Bottom) relative
	    populations of the clusters. (a-c) DBM trained on sPS liquids
	    using (a) prior $p_1$, (b) prior $p_2$, (c) no prior. (d-f) DBM
	    trained on octane and cumene liquids using (d) prior $p_1$, (e)
	    prior $p_2$, (f) no prior. ``ol'' refers to outlier.}
	    \label{hm}
	\end{center}
\end{figure*}

An in-depth analysis of the low-dimensional maps is shown in
Fig.~\ref{hm}. We identify cluster centers using a $k$-means
algorithm. We then assign each point of the two dimensional projection
to its closest cluster. This allows us to compute a confusion matrix
comparing the cluster assignments of reference and backmapped
structures. The diagonal on the confusion matrix hereby refers to reference and backmapped structures that get mapped into the same cluster, indicating closeness in conformational space. The results for the chemically-specific and
chemically-transferred models can be found in Fig.~\ref{hm} (a--c) and
Fig.~\ref{hm} (d--f), respectively. Interestingly, the confusion
matrix becomes most diagonal for both training sets if we train DBM
without any prior (Fig.~\ref{hm} c, f). However, the reduced CG resolution implies that a single CG
configuration will correspond to an ensemble of atomistic microstates. This ensemble might span a broad region in conformational space and two microstates corresponding to the same CG structure do not necessarily have to fall into the same cluster. Therefore, it is not clear to
what extent backmapped structures should correctly map to the same
clusters as their corresponding reference structures. More importantly, the relative clusters populations should
match, as this would indicate an accurate coverage of the
conformational space. Regarding models trained on sPS, we find that
both prior $p_2$ and no prior lead to an excellent match of the
relative populations. For models trained on octane and cumene all
priors lead to comparable accuracy, with limited reproduction of the
relative populations.

\section{Conclusion}
\label{sec:conclusion}

In this study we probe the chemical transferability of our machine
learning (ML) model deepBackmap (DBM), designed for backmapping of
condensed-phase molecular structures. To this end, we trained DBM
solely on liquid-phase configurations of small molecules, specifically
octane and cumene, and then tested its performance on the more complex
system of syndiotactic polystyrene (sPS) melts. Furthermore, we tested
different priors and their impact on the quality of the backmapped
structures.

We observe that the local correlations learned from the
chemically-transferred (i.e., octane and cumene) transfer remarkably
well to sPS. Despite discrepancies for some structural distributions,
such as the angles of the backbone carbon atoms, the overall quality
of the backmapped structures is encouraging. Importantly, the model
performs well in a challenging condensed-phase environment, and is
able to reproduce the distribution of Lennard-Jones energies with high
accuracy. Non-bonded structural features, in particular the pair
correlation function, match the reference distributions virtually
identically. A higher-order investigation, facilitated by the
Sketchmap algorithm, also indicates high structural fidelity. Although backmapped structures are not necessarily mapped onto the same cluster as their corresponding reference structures, as shown by the confusion matrices, DBM is
able to cover the correct spots in conformational space. The relative statistical weight
of generated microstates leaves further room for improvement.

The results shown here indicate that a sequential reconstruction
combined with a local-environment representation are well suited
toward chemical transferability. However, generalization shows its
limits. For example, the orientation of the phenyl ring with respect
to the backbone cannot be learned from the octane and cumene
structures, leading to misplaced atoms. This likely explains the
limited quality of the carbon-backbone structures
(Fig.~\ref{results_angles} a,b and Fig.~\ref{results_dihs} a,b). In
addition, force-field inconsistencies between molecules will evidently
lead to incoherent conformational spaces, directly affecting the
transferability of the backmapping.

We investigated the role of the prior. The different
priors only have a marginal impact on the quality of the covalent
interaction terms. On the contrary, the non-bonded Lennard-Jones interaction is more
sensitive to the prior, as can be seen in the distributions of
Lennard-Jones energies. This can be explained with the functional form of the interactions: While the harmonic or periodic potentials for the bonded interactions react moderately to shifts of the atomic arrangement, the Lennard-Jones potential is more sensitive and small shifts can rapidly change the energy by several orders of magnitude. The energy-minimizing prior $p_1$ leads to
high-quality configurations for the chemically-transferred DBM, but
yields too low energies when trained directly on sPS. The
energy-matching prior $p_2$ has an overall negligible impact compared
to training without any prior. We believe that prior $p_1$ encourages
the model to learn more general aspects, such as increasing the distance
of non-bonded atoms, while prior $p_2$ and no prior (only
data-driven) let the model focus on more specific features, making
them less generalizable.

In general, our approach offers the perspective to efficiently recycle
learned local correlations from small and easy to sample molecules and
deploy them for the backmapping of more complex systems. This can be
of tremendous use for generating high resolution configurations of
complex systems, without necessarily simulating the fine-grained
system first.

\section*{Acknowledgments}

The authors thank Yasemin Bozkurt Varolg\"une\c{s} and Arghya Dutta
for critical reading of the manuscript. We are grateful to Chan Liu
for providing coarse-grained and atomistic simulations of syndiotactic
polystyrene. This work was supported in part by the SFB-TRR146
Collaborative Research Center ``Multiscale Simulation Methods for
Soft Matter” of the Deutsche Forschungsgemeinschaft, as well as the Max
Planck Graduate Center. 

\section*{Data Availability Statement}

The data that support the findings of this study are available from
the corresponding author upon reasonable request.

\section*{References}
\bibliographystyle{rsc}
\bibliography{references}

\end{document}